# ChatLogo: A Large Language Model-Driven Hybrid Natural-Programming Language Interface for Agent-based Modeling and Programming

ChatLogo: An LLM-Driven Interface for Agent-based Modeling and Programming

John Chen

Northwestern University, civitas@u.northwestern.edu

Uri Wilensky

Northwestern University, uri@northwestern.edu

Building on Papert (1980)'s idea of children talking to computers, we propose ChatLogo, a hybrid natural-programming language interface for agent-based modeling and programming. We build upon previous efforts to scaffold ABM & P learning and recent development in leveraging large language models (LLMs) to support learning of computational programming. ChatLogo aims to support conversations with computers in a mix of natural and programming languages, provide a more user-friendly interface for novice learners, and keep the technical system from over-reliance on any single LLM. We introduced the main elements of our design: an intelligent command center, and a conversational interface to support creative expression. We discussed the presentation format and future work. Responding to the challenges of supporting open-ended constructionist learning of ABM & P and leveraging LLMs for educational purposes, we contribute to the field by proposing the first constructionist LLM-driven interface to support computational and complex systems thinking.

## 1 Introduction

In Mindstorms, Seymour Papert's pioneering book on Constructionism, a central motif was to support children talking to computers. Instead of using computers to "program" children, children gain control of computers by programming them. Consequently, the Logo programming language family opens vast possibilities for learning in mathematics (e.g. through Logo, [10]), in physics (e.g. through DynaTurtle [4]), as well as in complex systems (e.g. through NetLogo [16]). Like the original Logo language, to empower children in learning to "talk to computers", designers of Logo descendants strive to make their syntax close to natural languages. Whereas, programming languages, however close to natural forms of talking, still require a formal system of syntax and vocabulary.

In this proposal, we focus on NetLogo [16], the most widely used programming language for agent-based modeling and programming (ABM & P) in the Logo family. Agent-based modeling (ABM) is a powerful methodology that leverages simple computational rules for individual agents to produce complex emergent phenomena [17]. Agent-based programming (ABP) is a decentralized and often probabilistic programming paradigm that serves as the technical foundation of ABM [2]. While ABM has been widely employed in educational settings, facilitating deep engagement with ABM still poses challenges for teachers and learners, partly due to NetLogo's formal structures and vocabulary, and partly due to ABP being a different paradigm than what is usually taught at school [2].

While many efforts have been done to scaffold the learning of ABM & P, only a number of them are dedicated to open-ended learning contexts (e.g. [11] [3]). Meanwhile, recent advances in large language models (LLMs), have opened up new opportunities for supporting open-ended constructionist learning of NetLogo. While not directly evaluated on NetLogo, codex, GPT-3.5, and GPT-4 have all demonstrated considerable performance in general programming tasks. With their recent usage in education [9], it seems that "talking to computers" in a natural language context finally comes within reach. Building on those recent efforts, we present the design of ChatLogo, an LLM-driven hybrid natural-programming language interface for agent-based modeling and programming.

## 2 Background

ChatLogo is inspired by two lines of previous literature: efforts to support constructionist learning of ABM & P; advances in LLMs and conversational programming interfaces.

While a constructionist learning approach of ABM would naturally entail ABP to support learners' exploration, modification, and creation of agent-based models, many previous implementations stop short of coding in NetLogo (e.g. [5]). As ABMs are often integrated into science or social science curricula, programming often incurs a higher overhead for teaching and learning, since teachers and students are less prepared for the CS-related content [12]. Responding to this challenge, several studies tried to create block-based programming interfaces for NetLogo (e.g. [6]). While such interfaces could get children to start coding in 1-2 minutes [7], a trade-off always exists between the "floor" and "ceiling": the threshold for initial engagement, and the potential for expression [3]. As the power of block-based interfaces increases, they start to ask for scaffolding as well. For example, our recent study [3] found that interactive scaffolds significantly increased online young learners' short-term and long-term engagement with a block-based ABP environment. Pluralism was identified as a key element that contributed to the improvement: with several scripted pathways, the conversational experience for learners to build their own projects encouraged them to come back again.

However, there is always a limitation for pre-scripted scaffolds, as they became less efficient when young learners came up with their own project ideas [3]. The advent of advanced LLMs brought new hopes. Compared to earlier attempts at conversational programming interfaces that are still syntactically constrained (e.g. [15]), state-of-art LLMs such as GPT, PaLM, or LLaMA are capable of handling much more flexible or even malformed human inputs and translating them into programming languages (e.g. [13]) A few pioneering studies have been conducted to evaluate the effectiveness of LLMs in supporting the learning of programming languages. For example, [9] designed a Codex-powered interface and found short-term learning benefits for novice programmers. While promising, LLMs also come with limitations: they are prone to mistakes, hallucinations, potential biases, or harmful language. [14] found that professional programmers' task completion rates or time were not improved by GitHub Copilot, partly because participants felt difficulty in understanding and debugging generated code. [8] found that participants felt they must learn the LLMs' "syntaxes" and struggled to form an accurate mental model to interact with LLMs. They also performed worse in domain-specific tasks, e.g. in NetLogo.

## 3 Design goals

ChatLogo is designed as a web-based system with three goals in mind:

1. **Support novice programmers to "talk to computers" in a mix of programming and natural languages.** Both Logo and NetLogo are implicitly conversational. By placing a "command center" in parallel to the main view, the user would communicate with the computer through text messages or changes in the view. However, there are always correct ways to talk to computers, which take time for learners to grasp. Our design needs to bridge the gap between natural and programming languages by accepting both of them and talking back to learners in a more natural way.
2. **Provide a more friendly interface for learners with no or little computer science backgrounds to creatively express themselves by programming computers.** Even with the latest LLM-based interfaces, learners still struggled to find out the "correct" way to interact with computers [8]. LLMs also frequently provide incorrect responses that require expertise in computer science to identify and resolve. Consequently, LLM-based interfaces are currently more beneficial for learners with more prior programming experiences [9]. While eliminating the underlying issues of LLMs are beyond our means, our design should tailor the system for novice learners - rather than tailor novice learners for LLMs.

3. **Keep the technical system from over-reliance on any single LLM.** We recognize the inherent risk in relying on a private-owned LLM. For example, many studies cited in this paper leveraged OpenAI's Codex model released in 2022. Within a year, OpenAI would shut down public access to the model, making replications of those latest studies all but impossible if not for a selected few. There are also fresh and valid concerns about data privacy, especially when children and schools could be potential users of our design. To mitigate this risk, we intentionally build our system on a less powerful general-purpose LLM (gpt-3.5-turbo instead of gpt-4) and ensure that the design would eventually work with other (fine-tuned) LLMs that could eventually be deployed in a local and secure environment.

## 4 Design overview

We briefly describe the prototype design of ChatLogo, a hybrid natural-programming language interface for agent-based modeling and programming. A web-based browser-server system, ChatLogo is built with both LLMs and conventional programming. It is highly modularized: the underlying LLM could be replaced at no cost, and its features could be selectively enabled or disabled depending on the learning needs. The system could be adapted for other languages as well.

### 4.1 An Intelligent Command Center

ChatLogo is an intelligent command center of NetLogo. In this example, we showcase a classical mistake of novice NetLogo programmers: try to `set color` of patches. In NetLogo desktop's command center (Appendix 1), the input box would deny the entrance of such an ill-formatted input and show an error message instead. It is as if the computer tells the user back: The way you talked was wrong. I will not respond until you figure out the correct way. In Turtle Universe, the mobile incarnation of NetLogo [1], we made a slight improvement by introducing the help feature: in Appendix 2, the computer briefly explains the primitive and suggest some alternatives. However, it still requires the user to initiate the action, and we found relatively few users would touch the "Help" button [3].

At a surface level, ChatLogo inherited this interactive design. However, its behavior diverges when the user gave a malformed NetLogo input (Appendix 3): besides an error message, it further provides two AI-driven options that could explain the error messages or fix the code. Appendix 4 demonstrates the explanation pathway. Once the AI finishes the answer, the learner could ask a follow-up question in natural language, or ask the AI to fix it for them. At this point, the AI would stress that it might make more mistakes: instead of taking away the learners' initiative, learners are still in charge of the loop. Alternatively, if they decide to send in a new NetLogo command instead, ChatLogo would attempt to execute it directly.

### 4.2 A Conversational Interface for Creative Expression

An intelligent command center might serve novice learners of NetLogo better. However, it assumes that the learner already knows something about the language, or the input would become unrecognizable in the eyes of the NetLogo compiler. A novice learner might talk in a more "conversational" way: I want to change the background color to red; or, I want to make turtles move around; or more broadly, I want to create a game of ants. A younger learner might also make spelling mistakes along the way, negatively affecting LLMs' performance. We further notice that: especially for LLMs trained to be a chatbot (e.g. gpt-3.5-turbo or gpt-4), they tend to give a long answer for most questions and make decisions for the learner before asking for clarification. For example, Appendix 5 demonstrates GPT-4's answer to a simple question: "In NetLogo, how can I create some moving turtles?" Its answer not only assumed much on the learner's behalf, e.g., turtles would turn back 180 degrees when hitting the edge of the world; it gave the learner step-by-step instructions to follow. In a way, GPT-4 attempts to program the learner.

Our approach differs from the pre-trained GPT-4 behavior (Appendix 5). Instead of right away writing code and giving instructions, ChatLogo attempts to clarify the learners' needs and intention (Appendix 6). Instead of sending large chunks of code directly to the learner, it attempts to co-develop the NetLogo code. As shown in Appendix 7, the learner is free to edit the code: either in NetLogo, or in natural languages through the "Ask" feature. Instead of overclaim the correctness of the code, it admits the possibility of making mistakes, and co-works with the learner to address the potential issues (Appendix 8). Finally, upcoming features of ChatLogo will allow learners to add the human-AI co-created code back to the NetLogo model and help learners plan out entire projects in their mind.

## 5 FUTURE WORK

Despite its potential, there is still a long way to go before ChatLogo could be safely and effectively deployed to K-12 educational settings. More work needs to be done to reduce its mistakes, hallucinations, and potentially harmful language. As we do not expect LLMs to solve these fundamental problems overnight, we are also interested in understanding how human-computer interaction and learning design could be leveraged to mitigate the potential harm and develop learners' AI literacy along the way. To achieve this, we are currently running a study with adult NetLogo programmers and evaluating if it would be appropriate to work with children.

There has been much debate around LLMs and the future of humanity as of late. Our ultimate hope is that LLMs could become a liberating force, instead of an oppression one, for both children and adults. This requires children to be able to program computers for their own purposes, not vice versa. This asks for a more constructionist future for education, where children could be better equipped and supported to construct their own meaningful artifacts, not vice versa.

# A APPENDICES

## A.1 Screenshot of NetLogo Desktop's Command Center

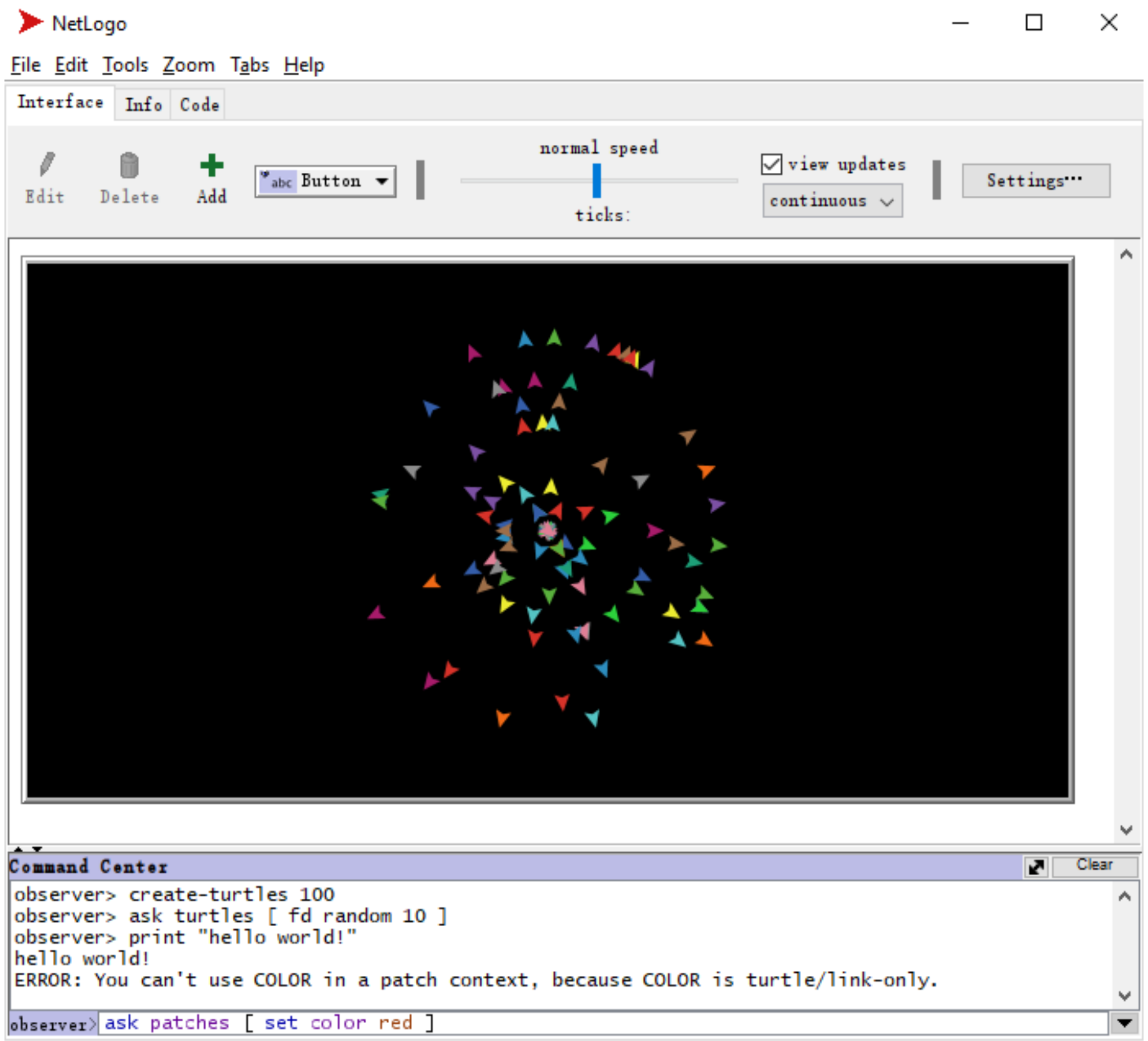

### A.2 Screenshot of Turtle Universe's Command Center

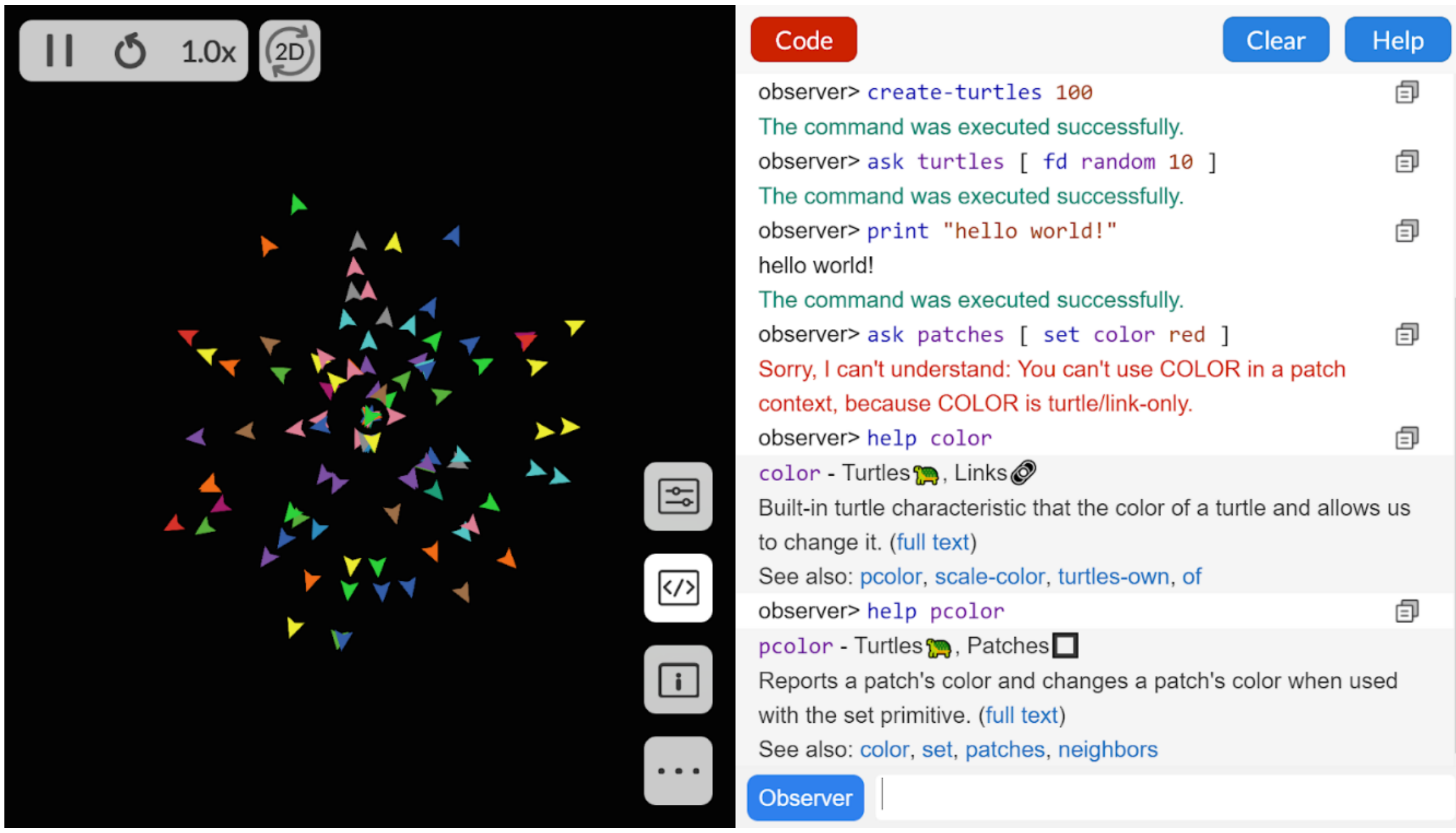

### A.3 Screenshot of ChatLogo running on Turtle Universe

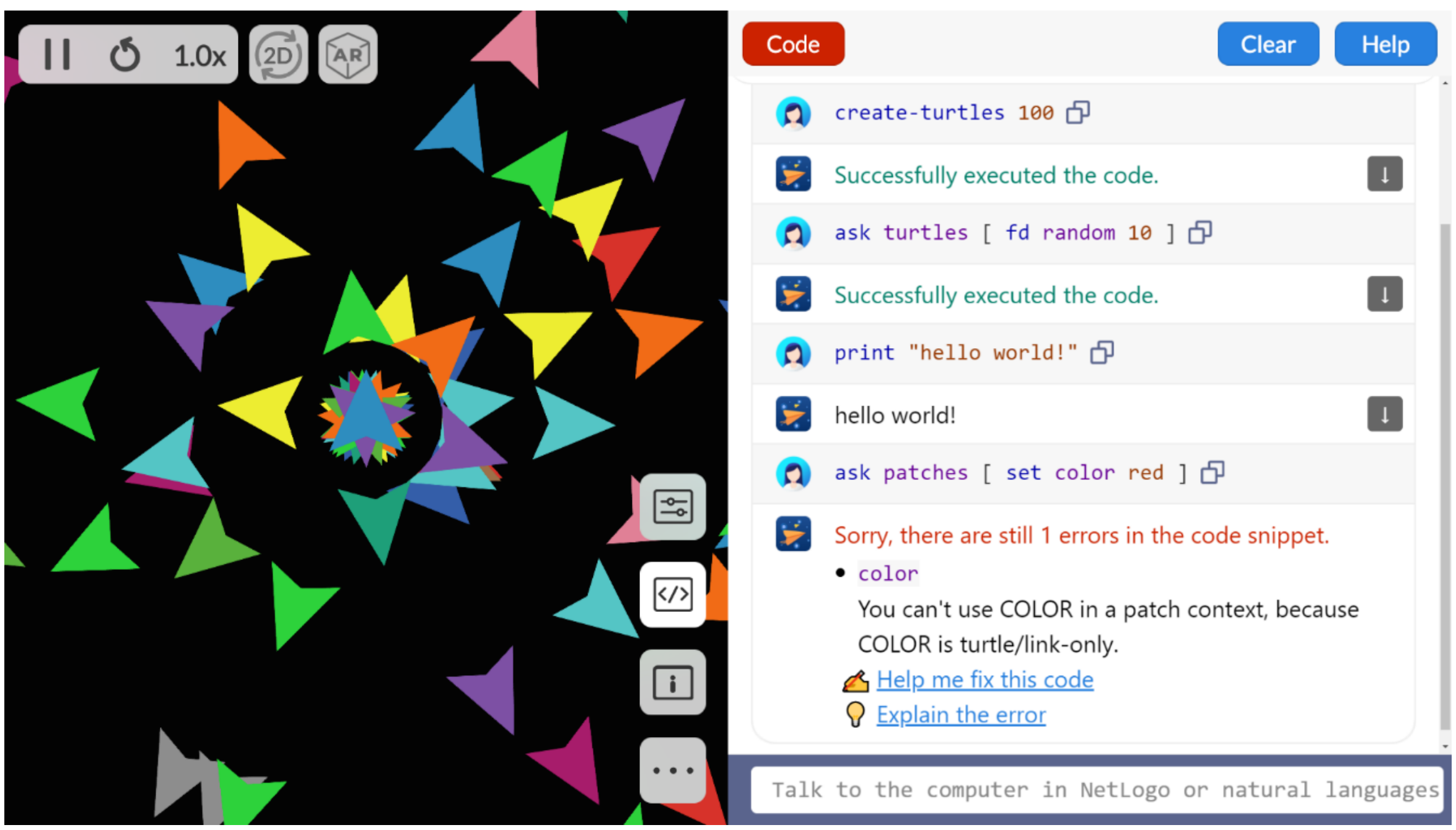

### A.4 Screenshot of ChatLogo running on Turtle Universe, Continued

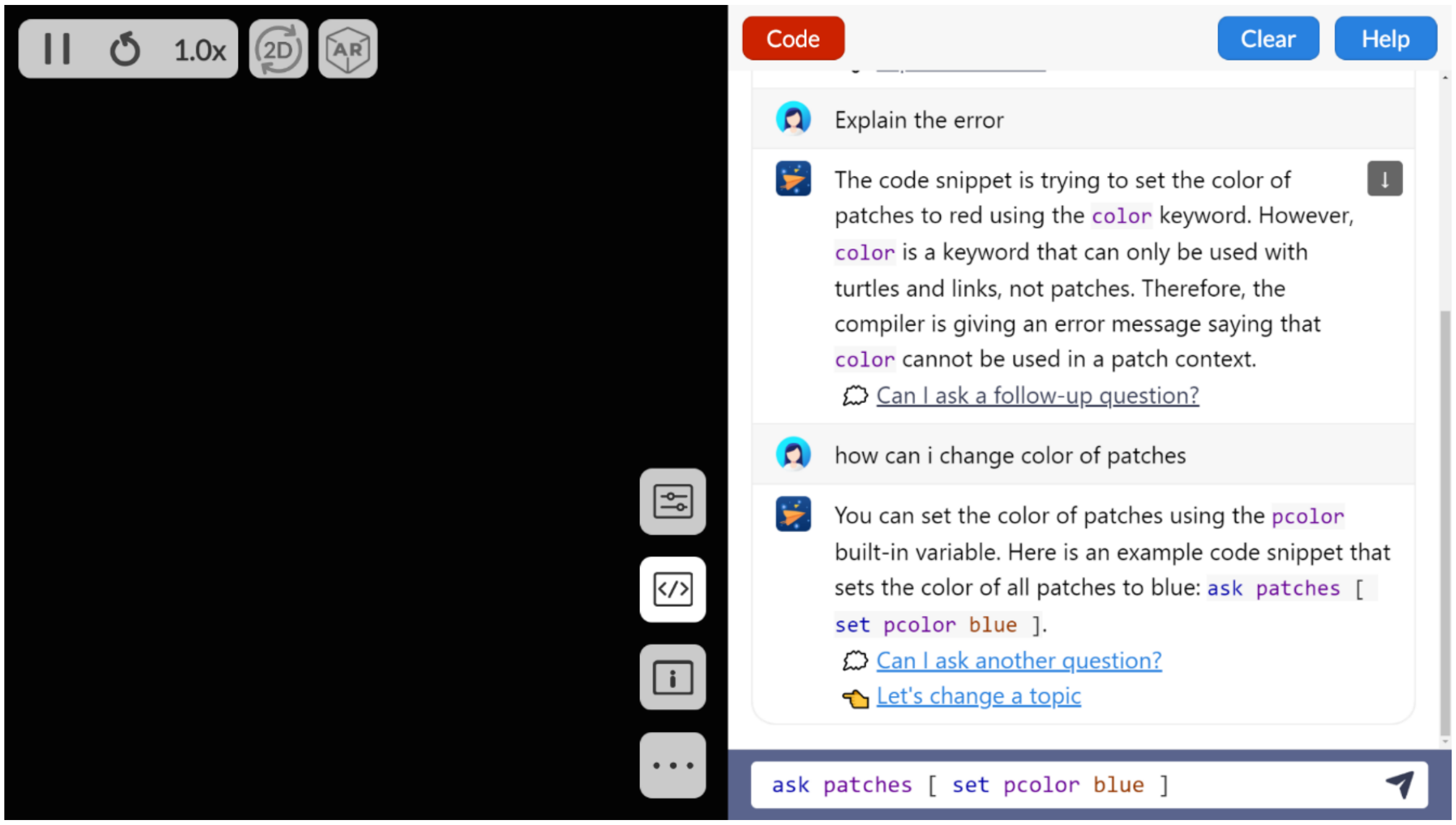

### A.5 GPT-4's response to "create some moving turtles"

**USER** In NetLogo, how can I create some moving turtles?

**ASSISTANT** In NetLogo, you can create moving turtles by defining a set of turtle properties and behaviors, and then using the `ask` command to update their positions in each tick of the simulation. Here's a simple example of how to create moving turtles in NetLogo:

1. Create a new NetLogo model or open an existing one.

2. In the "Code" tab, define a new breed of turtles called "moving-turtles":

```
breed [moving-turtles moving-turtle]
```

3. Define a procedure to set up the model. In this example, we'll create 10 moving turtles and set their initial properties:

### A.6 ChatLogo's response to "create some moving turtles"

create moving turtles

It seems that you have several different needs. Let's do one at a time. Which one do you want to start with?

- Create turtles
- Make turtles move
- Let me clarify it
- Let's change a topic

Create turtles

Working on: **create turtles**
Sure, I can help you with that. Can you please provide me with more information?

What do you want to call the turtles in the code?

turtles

e.g. turtles rabbits cars

How many turtles do you want to create?

10

e.g. 10 50 random between 20-30

Where do you want to create the turtles?

random

e.g. random at (0,0) around a specific patch

### A.7 The First Iteration of Code Provided by ChatLogo

```
; Create 10 turtles using the breed name "turtles"
create-turtles 10 [
  ; Set the turtles" positions randomly
  setxy random-xcor random-ycor
]
```

Run ? Ask 1 / 1

Below is a summary of my request:

- breed: turtles
- number: 10
- position: random

I am working on a first version of the code. The code might have mistakes.

- Let's change a topic

### A.8 Ask ChatLogo to Fix a Simple Bug Made by the Researcher

```
; Move all turtles
ask turtle [
  ; Set heading to up
  set heading 90
  ; Move forward random between 1-2 units
  fd (1 + random 2)
]
```

Run ? Ask Back 3 / 3

Trying to run the code...

Sorry, but we need to fix the 1 errors in the code (marked with __red squiggly lines__) before continuing.

Help me fix this code

Help me fix this code

Sure, I am working on the fixed code. Note that the code can still have mistakes.

```
; Revised code and line comments and explanations
; Move all turtles
ask turtles [
  ; Set heading to up
  set heading 90
  ; Move forward random between 1-2 units
  fd (1 + random 2)
]
```